\RequirePackage{fix-cm}
\documentclass[smallextended]{svjour3}       
\smartqed  
\usepackage{graphicx}
\usepackage{hyperref}       
\usepackage{url}            
\usepackage{booktabs}       
\usepackage{amsfonts}       
\usepackage{nicefrac}       
\usepackage{microtype}      
\usepackage{lipsum}
\usepackage{amsmath,amssymb,dsfont,mathrsfs,bbm}
\usepackage{mathtools,multirow,xfrac}
\usepackage{subcaption}
\usepackage[justification=centering]{caption}
\usepackage{xcolor}
\usepackage{chemfig}
\usepackage[utf8]{inputenc}
\usepackage[linesnumbered,ruled,vlined]{algorithm2e}
\SetKwInput{KwInput}{Input}                
\SetKwInput{KwOutput}{Output}

\begin{document}

\title{Large independent sets \\ on random $d$-regular graphs with fixed degree $d$}


\author{Raffaele Marino          \and
        Scott Kirkpatrick  
}

\institute{ Raffaele Marino  \at
  Laboratoire de Th\'eorie des Communications,\\
  Facult\'e Informatique et Communications,\\
  \'Ecole Polytechnique F\'ed\'erale de Lausanne,\\
  1015, Lausanne, Switzerland\\
  Dipartimento di Fisica, Sapienza Università di Roma,\\
  P.le Aldo Moro 5, Roma, 00185, Italy\\
   \texttt{raffaele.marino@uniroma1.it}\\
           \and
Scott Kirkpatrick \at
    School of Computer Science and Engineering,\\
    The Hebrew University of Jerusalem,\\
    Edmond Safra Campus, Givat Ram, Jerusalem 91904, Israel\\
}

\date{Received: date / Accepted: date}

\maketitle

\begin{abstract}

This paper presents a linear prioritized local algorithm that computes large independent sets on a random $d$-regular graph with  small and fixed degree $d$. We studied experimentally the independence ratio obtained by the algorithm when $ d \in [3,100]$. For all  $d \in [5,100]$, our results  are larger than lower bounds calculated by exact methods, thus providing improved estimates of lower bounds. 


\end{abstract}

\section{Introduction}
\label{intro}
Given a graph $G(\mathcal{N}, E)$, where $\mathcal{N}$ is the set of vertices of cardinality $|\mathcal{N}|=N$ and $E$ the set of edges of cardinality $|E|=M$, finding the maximum set of sites no two of which are adjacent is a very difficult task. This problem is known as the maximum independent set problem (MIS). It was shown to be NP-hard, and no known polynomial algorithm can guarantee to solve it \cite{cook2006p}. In other words, finding a set $\mathcal{I}$ of vertices, with the maximum cardinality, such that for every two vertices $i,j \in \mathcal{I}$, there is no edge connecting the two, i.e., $(i,j) \notin E$, needs a time which is super-polynomial if $P\neq NP$. 

For example, the first nontrivial exact algorithm for the MIS was due to Tarjan and Trojanowski’s $O(2^{N/3}) \sim O(1.2599^N)$ algorithm in 1977 \cite{tarjan1977finding}. Since then, many improvements have been obtained.  Today, the best algorithm that can solve the MIS exactly needs a time $O(1.1996^N)$ \cite{xiao2017exact}. Those results are a worst case bound.   We direct the interested reader to  \cite{xiao2017exact}, and references therein, for a complete discussion on exact algorithms. 

The MIS  is important for applications in Computer Science, Operations Research, and Engineering, such as graph coloring, assigning channels to the radio stations, register allocation in a compiler, etc. 

Besides having several direct applications \cite{bomze1999maximum}, the MIS is closely related to another well-known optimization problem, the maximum clique problem \cite{karp1972reducibility},\cite{marino2018revisiting}. For finding the maximum clique (the largest complete subgraph) of a graph $G(\mathcal{N}, E)$, it suffices to search for the maximum independent set of the complement of $G(\mathcal{N}, E)$.

The MIS has been studied on many different random structures, like Erd\H{o}s-Rényi graphs (ER),  random $d$-regular graphs (RRG). A random $d$-regular graph  is a graph selected from the distribution of all $d$-regular graphs on $N$ vertices, with $Nd$ even. A regular graph, is defined as a graph where each vertex has the same number of neighbors, i.e., $d$. Random $d$-regular graphs represent a subset of  Erd\H{o}s-Rényi graphs distribution with probability $p=<d>/N$.

For the Erd\H{o}s-Rényi class $G_{ER}(N,p)$, where $p$ is the probability that two different vertices are connected to each other, known local-search algorithms can find solutions only up to half the maximum independent set present, which is $\sim 2 \log_{1/(1-p)} N$ \cite{wein2020optimal} in the limit $N\to \infty$. 

This behavior also appears for random $d$-regular graphs $G_d(N)$. 
In this cases, for example, Gamarnik and Sudan \cite{gamarnik2014limits} showed that, for a sufficiently large value of $d$, local algorithms cannot find the size of the largest independent set in a $d$-regular graph of large girth with an arbitrarily small multiplicative error. 



The results of Gamarnik and Sudan \cite{gamarnik2014limits}  was successively improved by Rahman, and Virág \cite{rahman2017local}, which analyzed the intersection densities of many independent sets in random $d$-regular graphs. They proved that for any $\epsilon > 0$, local algorithms cannot find independent sets in random $d$-regular graphs with an independence ratio larger than $(1 + \epsilon) \frac{\ln d}{d}$ if $d$ is sufficiently large. The independence ratio is defined as the density of the independent set, thus $\alpha=|\mathcal{I}|/|\mathcal{N}|$. Recently, the exact value of the independence ratio for all sufficiently large $d$ was given by Ding et al. \cite{ding2016maximum}.


However, these results appear to say nothing about small and fixed $d$. When $d$ is small and fixed, e.g., $d=3$ or $d=30$, indeed, only lower and upper limits, expressed in terms of independence ratio, are known. 

Lower bounds on the independent sets' size identify sets that an efficient algorithm can find, while upper bounds are on the actual maximum independent set, not just on the size an algorithm can find.


The first upper bound for such a problem was given in 1981 by Bollob\'as \cite{bollobas1981independence}. He showed that the supremum of the independence ratio of $3$-regular graphs with large girth is less than $6/13 \sim 0.461538$, in the limit $N \to \infty$. 

McKay, 1987, improved and generalized this result to $d$-regular graphs with large girth \cite{mckay1987}, by using the same technique and a much more careful calculation. For example, for the cubic graph ($3$-regular graph), he was able to push Bollob\'as upper bound down to $0.455370$. However, since then, only for cubic graphs, the upper bound has been improved by Balogh et al. \cite{balogh2017cubic}, namely to $0.454$.  Replica methods suggest a slightly lower upper bound, and thus a smaller gap at small values of $d$ \cite{barbier2013hard}. For examples, the upper bounds given in  \cite{barbier2013hard} for $d=3$ is $0.4509$, while for $d=4$ is $0.4112$.  A recent paper shows that this approach can be proven, but again, only for large $d$\cite{ding2016maximum}.  

Remarkable results for lower bounds were obtained first by Wormald in 1995 \cite{wormald1995differential}. He considered processes in which random graphs are labeled as they are generated and derived conditions under which parameters of the process concentrate around the values of real variables which come from the solution of an associated system of differential equations. By solving the differential equations he computed lower bounds for any fixed $d$ returned by a prioritized algorithm, improving the values of bounds given by Shearer \cite{shearer1983note}.


This algorithm is called \textit{prioritized} because there is a priority in choosing vertices added to the independent set \cite{wormald2003analysis}. It follows the procedure of choosing vertices in the independent set $\mathcal{I}$ one by one, with the condition that the next vertex is chosen randomly from those with the maximum number of neighbors adjacent to vertices already in $\mathcal{I}$.
	After each new vertex in $\mathcal{I}$ is chosen (or labeled with an $I$), we must complete all of its remaining connections and label the neighbors which are identified as members of the set $\mathcal{V}$ (for vertex cover). Although each vertex in $\mathcal{I}$ can be chosen according to its priority, the covering vertices that complete its unfilled connections must then be chosen at random amount the remaining connections, to satisfy Bolobas' configuration model \cite{wormald1995differential}. Following this priority is a simple way to minimize the size of the set of vertices covered and maximize the number of sites remaining as candidates for the set $\mathcal{I}$.

More precisely, we are given a random $d$-regular graph $G_d(N)$, and we randomly choose a site $i$ from the set of vertices $\mathcal{N}$. We set $i$ into $ \mathcal{I}$, and we set all the vertices neighboring $i$ into a set $\mathcal{V}$. We label elements of $\mathcal{I}$ with the letter $I$, while elements of $\mathcal{V}$ with the letter $V$. Then, from the subset of vertices in $\mathcal{N}$ that are neighbors of vertices in $\mathcal{V}$, but are not yet labeled $I$ or $V$, we choose randomly the element $k$ that has the maximum number of connections with sites in $\mathcal{V}$. We set it into $\mathcal{I}$. The vertices neighboring $k$, which are not in $\mathcal{V}$, are added to the set $\mathcal{V}$. This rule is repeated until  $|\mathcal{I}|+|\mathcal{V}|=N$. Along with this algorithm, one can consider an associated algorithm that simultaneously generates the random $d$-regular graph $G_d(N)$ and labels vertices with the letter $I$ or $V$. This associated algorithm, which will be described in detail in the next sections, allowed Wormald to build up the system of differential equations used for computing lower bounds for the MIS.

Improvements over this algorithm were achieved by Duckworth et al. \cite{duckworth2009large}. These improvements were obtained by observing, broadly speaking, that the size of the structure produced by the algorithm is almost the same for $d$-regular graphs of very large girth, as it is for a random $d$-regular graph. However, since then, new lower bounds have been achieved only at small values of $d$, e.g., $d=3$ and $d=4$. Interesting results at $d=3$ have been achieved by Csóka, Gerencsér, Harangi and Virág \cite{csoka2015invariant}. They were able to find an independent set of cardinality up to $0.436194 N$ using invariant Gaussian processes on the infinite $d$-regular tree. This result was once again improved by Csóka \cite{csoka2016independent} alone, which was able to increase the cardinality of the independent set on large-girth $3$-regular graph up to $0.445327 N$ and on large-girth 4-regular graph up to $0.404070 N$, by solving numerically the associated system of differential equations.
\begin{table}
\label{tab-res}
\begin{center}
\begin{tabular}{ |c|c|c|c| } 
\hline
$d$ & $\alpha_{UB}$ & $\alpha_{LB}$ & $\alpha_{\infty} \pm z_{99\%}\sigma_{\alpha_{\infty}}$\\
\hline
3 & 0.45400 & 0.44533 &  0.44533 (1) \\ 
4 & 0.41635 & 0.40407 & 0.40087 (2) \\ 
5 & 0.38443 & 0.35930 & $\mathbf{0.36476  (2)}$\\ 
6 & 0.35799 & 0.33296 & $\mathbf{0.33600 (2)}$ \\ 
7 & 0.33567 & 0.31068 & $\mathbf{0.31241 (2)}$\\ 
8 & 0.31652 & 0.28800 & $\mathbf{0.29255 (1)}$\\ 
9 & 0.29987 & 0.27160 & $\mathbf{0.27555 (2)}$\\ 
10 & 0.28521 & 0.25730 & $\mathbf{0.26079 (2)}$ \\ 
20 & 0.19732 & 0.17380 & $\mathbf{0.17550 (4)}$\\ 
50 & 0.11079 & 0.09510 & $\mathbf{0.09574 (2)}$\\
100 & 0.06787 & 0.05720 & $\mathbf{0.05754 (1)}$\\
\hline
\end{tabular}
\end{center}
\captionsetup{font=footnotesize,justification=justified}
\caption{The table shows the values of upper and lower bounds for the independence ratio for random $d$-regular graph with small and fixed value of $d$. $d$ is the degree of the random $d$-regular graph. $\alpha_{UB}$ column describes the upper bound computed by  McKay in \cite{mckay1987}, and only for $d=3$ by Balogh et al. \cite{balogh2017cubic}. Upper bounds identifies the actual maximum value of the independent set can be. $\alpha_{LB}$ column identifies the best density of independent set obtained in \cite{wormald1995differential}, \cite{duckworth2009large}, \cite{hoppen2018local}, \cite{csoka2016independent} . Lower bounds identify the size of independent sets that an efficient algorithm can find.
The last column, i.e. $\alpha_{\infty} \pm z_{99\%}\sigma_{\alpha_{\infty}}$, instead, identifies the confidence intervals of the conjectured new bounds at the $99\%$.}
\end{table}

These improvements were obtained by deferring the decision whether a site $i\in \mathcal{N}$ must be labeled with a letter $I$ or $V$. More precisely, he  requires that the sites for which a decision is deferred need additional (temporary) labels.  This means that counting the evolution of their populations, either through a differential equation or by experiment, gets more complicated.

Csóka \cite{csoka2016independent} was able to improve lower bounds only for $d=3$ and $d=4$. This paper aims to generalize his method for any $d \geq 5$, using an experimental approach. We recall in Tab. $1$ the best upper and lower bounds\footnote{Recently in \cite{angelini2019monte} has been presented a Monte Carlo method that can experimentally outperform any algorithm in finding a large independent set in random $d$-regular graphs, in a (using the words of the authors) \textit{" running time growing more than linearly in N"} \cite{angelini2019monte}. These authors conjectured lower bounds improvements only for $d=20$ and $d=100$, but with experimental results obtained on random $d$-regular graphs of order $N=5\cdot 10^4$.  However, in this work, we are interested in comparing our results with the ones given by the family of prioritized algorithm because we believe that a rigorous analysis of the computational complexity would be performed on this algorithm.} for $ d\in [3,100]$, first and second columns respectively. 

In this paper, as stated above, we  present experimental results of a greedy algorithm, built upon existing heuristic strategies, which leads to improvements on known lower bounds of large independent set in random $d$-regular graphs $\forall d \in [5, 100]$ \cite{wormald1995differential}, \cite{hoppen2018local}, \cite{duckworth2009large}.

This new algorithm runs in linear time $O(N)$ and melds Wormald’s, Duckworth and Zito’s, and Csoka’s ideas of prioritized algorithms \cite{wormald1995differential}, \cite{duckworth2009large}, \cite{hoppen2018local}, \cite{csoka2016independent}. The results obtained here are conjectured new lower bounds for large independent set in random $d$-regular graphs.

They are obtained by inferring the asymptotic values that  our algorithm can reach when $N \to \infty$ and by averaging sufficient simulations to achieve confidence intervals at $99\%$. These results lead improvements on known lower bounds $\forall d \in [5, 100]$ that, as far as we know, are not reached by any other greedy algorithms. Although the gap with upper bounds is still present, these improvements may imply new rigorous results in finding a large independent set in random $d$-regular graphs.

The paper is structured as follows: in section  \ref{sec:definition}  we define our deferred decision algorithm, and introduce a site labelling which will identify those sites for which we defer the $I/V$ labelling decisions. In Sec. \ref{sec:d=3} we present the deferred decision algorithm for $d=3$, and we introduce experimental results obtained on random $3$-regular graphs of sizes \footnote{We recall that the order of a graph $G(\mathcal{N}, E)$ is the cardinality of its vertices set $\mathcal{N}$, while the size of a graph $G(\mathcal{N}, E)$ is the cardinality of its edges set $E$.} up to $10^9$. In Sec. \ref{sec:d>3}, we present our deferred decision algorithm for $d>3$, and the experimental results associated with it, by extrapolation on random $d$-regular graphs with sizes up to $10^9$ (see fourth column Tab. $1$). 

\section{Notation and the general operations of the deferred decision algorithm}
\label{sec:definition}

\begin{figure}
    \centering
    \includegraphics[width=0.8\columnwidth]{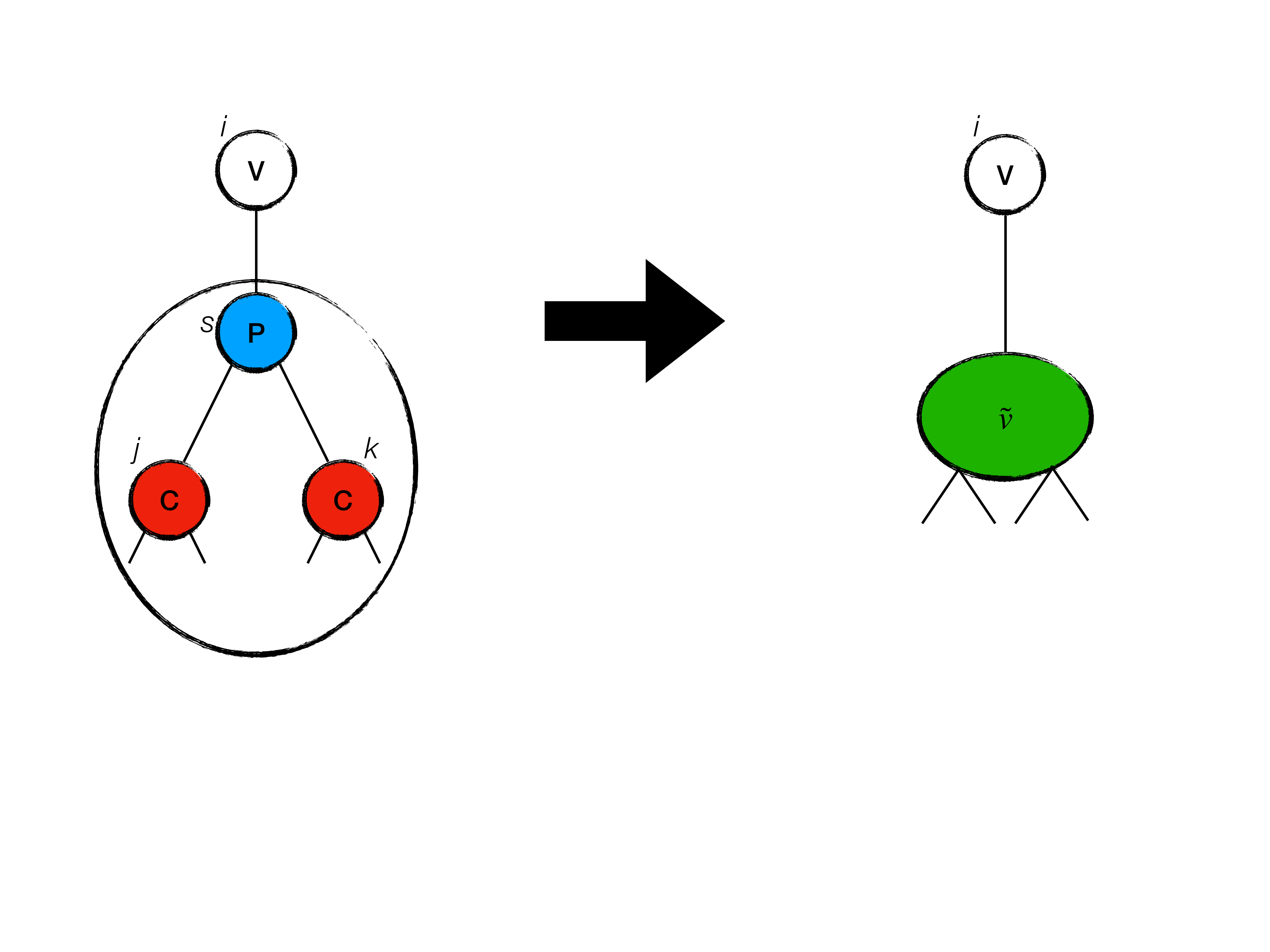}
    \captionsetup{font=footnotesize,justification=justified}
    \caption{\small The figure shows the equivalence of a $C-P-C$ structure with a $\tilde{v}$ site of anti-degree $\overline{\Delta}_{\tilde{v}}=4$. The site $s$, labeled $P$, is connected to site $i\in \mathcal{V}$, and with two random sites, $j$ and $k$  labeled $C$. The resulting structure is a virtual site $\tilde{v}\in {\mathcal{A}}$. } 
    \label{fig:compact}
\end{figure}

In this section, we define the notation used throughout this manuscript, and we define all operations that will be used to understand the deferred decision algorithm. 

We start by recalling that we deal with random $d$-regular graphs $G_d(N)$, where $d$ is the degree of each vertex $i\in \mathcal{N}$, where $\mathcal{N}$ is the set of vertices and $N=|\mathcal{N}|$.
All vertices $i \in \mathcal{N}$ are unlabeled.

For building a random $d$-regular graph we used the method described in \cite{wormald1995differential}, and introduced in \cite{bollobas1981independence}. 
\begin{definition}[\textbf{Generator of random $d$-regular graph Algorithm}]
\label{GA}
\textit{We take $dN$ points, with $dN$ even, and distribute them in $N$ urns labeled $1,2,\dots, N$, with $d$ points in each urn. We choose a random pairing $P=p_1, \dots, p_{dN/2}$ of the points such that $|p_i|=2 \forall i$. Each urn identifies a site in $\mathcal{N}$. Each point is in only one pair $p_i$, and no pair contains two points in the same urn. No two pairs contain four points from just two urns. For building a $d$-regular graph $G_d(N)$, then, we connect two distinct vertices $i$ and $j$ if some pair has a point in urn $i$ and one in urn $j$. The conditions on the pairing prevent the formation of loops and multiple edges. }
\end{definition}

The pairing referred to must be chosen uniformly at random, subjected to the constraints given. This can be done by repeatedly choosing an unpaired point and then choosing a partner for this point to create a new pair. As long as the partner is chosen uniformly at random from the remaining unpaired points, and as long as the process is restarted if a loop or multiple edge is created, the result is a random pairing of the required type \cite{wormald1995differential}.  

In this paper, we use the method above described so that while we generate the random $d$-regular graph $G_d(N)$, concurrently with our labelling process, labelling sites as we identify new links. 

The graphs built using the \textbf{Generator of random $d$-regular graph Algorithm} prevent the formation of loops and multiple edges, without introducing bias in the distribution where we sampling the graphs. We perform our analysis only on such graphs. In the case we meet on the last pair of vertices a loop, we set those sites not to be in the independent set. 

We define two separate sets $\mathcal{I}$ and $\mathcal{V}$ for independent and vertex cover sites. $\mathcal{I}$ identifies the set of graph nodes that satisfies the property that no two of which are adjacent and $\mathcal{V}$ its complement. A site $i \in \mathcal{I}$ is labeled with the letter $I$, while a site $j \in \mathcal{V}$, it is labeled with the letter $V$.

We define $\Delta_i$ to be the degree of a vertex $i$, i.e. the number of links that a site is connected with, while with $\overline{\Delta}_i$ the \textit{anti}-degree of a vertex $i$, i.e. the number of free connections that $i$ needs to complete during the graph building process. Of course, the constraint $\Delta_i + \overline{\Delta}_i = d$ is always preserved $\forall i \in \mathcal{N}$. At the beginning of the graph building process all $ i \in \mathcal{N}$ have $\Delta_i=0,\, \overline{\Delta}_i=d$. At the end of graph building process all graph nodes will have $\Delta_i=d,\, \overline{\Delta}_i=0$. We define $\partial i$ to be the set that contains all neighbors of $i$. 

For the sake of clarity, we define a simple subroutine on a single site $i\in \mathcal{N}$ of the \textbf{Generator of random $d$-regular graph Algorithm} (Subroutine GA($i$, $\overline{\Delta}_i$)) that will be useful for understanding the algorithm presented in the next sections. The Subroutine GA($i$, $\overline{\Delta}_i$) generates the remaining $\overline{\Delta}_i$ connections of site $i$.  It keeps the supporting data to reflect the evolution of the network growth.

 \begin{algorithm}[H]
 \label{algo1}

\KwInput{$i\in \mathcal{N}$, $\overline{\Delta}_i$;}
\KwOutput{$i$ connected with $d$ sites;}
Using the rules in \textbf{Definition \ref{GA} }, $i$ is connected randomly with $\overline{\Delta}_i$-sites; \\
$\overline{\Delta}_i=0$;\\
$\Delta_i=d$;\\
 \Return $i$ connected with $d$ sites;
 \caption{Subroutine GA($i$, $\overline{\Delta}_i$)}
\end{algorithm}

The sites that we choose following some priority, either the one we describe or any other scheme, will be called $P$ sites.  The sites which are found by following links from the $P$ sites (or by randomly generating connections from the $P$ sites) are called $C$ sites. More precisely, each site $j \in \mathcal{N}$  we choose, not labeled yet with any letter ($C$ or $P$) s.t.  $\overline{\Delta}_j\leq 2$ and the random connection(s) present on $j$ are on site(s) in $\mathcal{V}$   is a $P$ site. The set ${\mathcal{P}}$ defines the set of $P$ sites. The set  ${\mathcal{P}}$ is kept in ascending order with respect to the $anti$-degree $\overline{\Delta}_i$ of each site $i\in{\mathcal{P}}$. 

In general a site $i$ labeled $P$ will be surrounded by two sites labeled $C$. Because the labeling of those sites is deferred, we call those $C-P-C$ structures \textit{virtual} sites. A single \textit{virtual} site, $\tilde{v}$ has an anti-degree $\overline{\Delta}_{\tilde{v}}$ equal  to the sum of all anti-degrees of sites $l$ that compose site $\tilde{v}$, i.e.  $\overline{\Delta}_{\tilde{v}}=\sum_{l \in \tilde{v}}\overline{\Delta}_l$.  The number of sites $l \in \tilde{v}$ is equal to the  cardinality of $|\tilde{v}|$. The degree of  $\tilde{v}$  is 
$\Delta_{\tilde{v}}=d \,|\tilde{v}|-\overline{\Delta}_{\tilde{v}}$. As an example, we show in Fig. \ref{fig:compact} the operation of how a virtual site is created from a site $s\in {\mathcal{P}}$ with $\overline{\Delta}_{s}=2$, $\Delta_{s}=1$,  and two sites $j,k \in \mathcal{N}$ with $\overline{\Delta}_{j}=3$, $\overline{\Delta}_{k}=3$ and $\Delta_{j}=0$, $\Delta_{k}=0$. Let's assume that a site $s \in \mathcal{N}$ s.t. $\overline{\Delta}_{s}=2$ exists. This is possible because a site $l\in \mathcal{V}$ is connected with it. This means that $s$ must be labeled $P$ and put into $ {\mathcal{P}}$. Let's run Subroutine GA($s$, $\overline{\Delta}_s$) on $s$, and assume that the $s$ connects with two neighbors $j,k \in \mathcal{N}$. Being $j,k \in \mathcal{N}$ connected to a $P$ site, they are labeled $C$. We, then, define $\tilde{v}=\{s,j,k\}$. This set is a virtual node $\tilde{v}$ with  $\overline{\Delta}_{\tilde{v}}=4 $. 



\begin{figure}
    \centering
    \includegraphics[width=0.8\columnwidth]{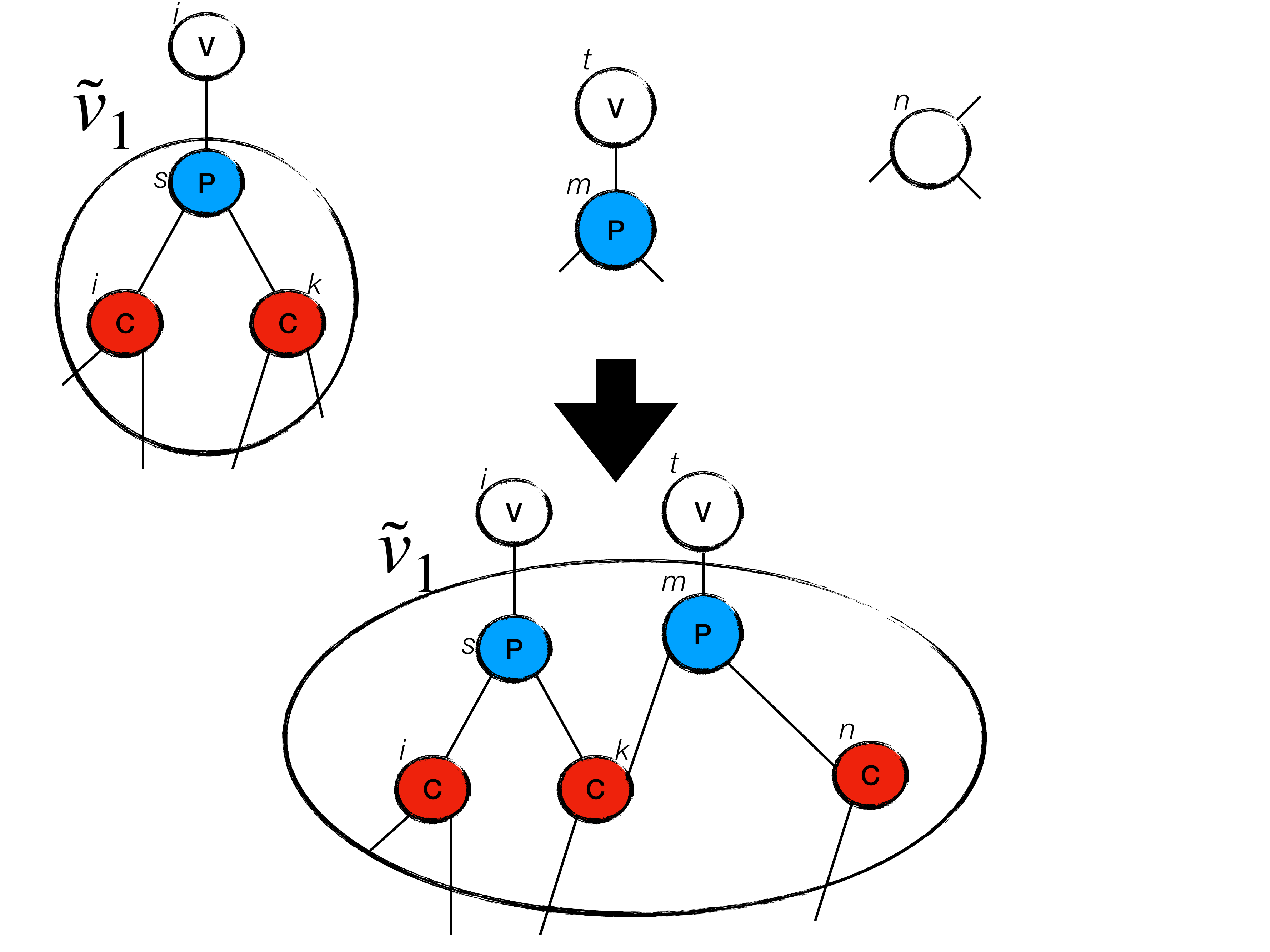}
    \captionsetup{font=footnotesize,justification=justified}
     \caption{\small The figure shows how the virtual site  ${\tilde{v}}_1\in {\mathcal{A}}$ is expanded. }
     \label{fig:chemic2}
 \end{figure}
\begin{figure}
     \centering
    \includegraphics[width=0.8\columnwidth]{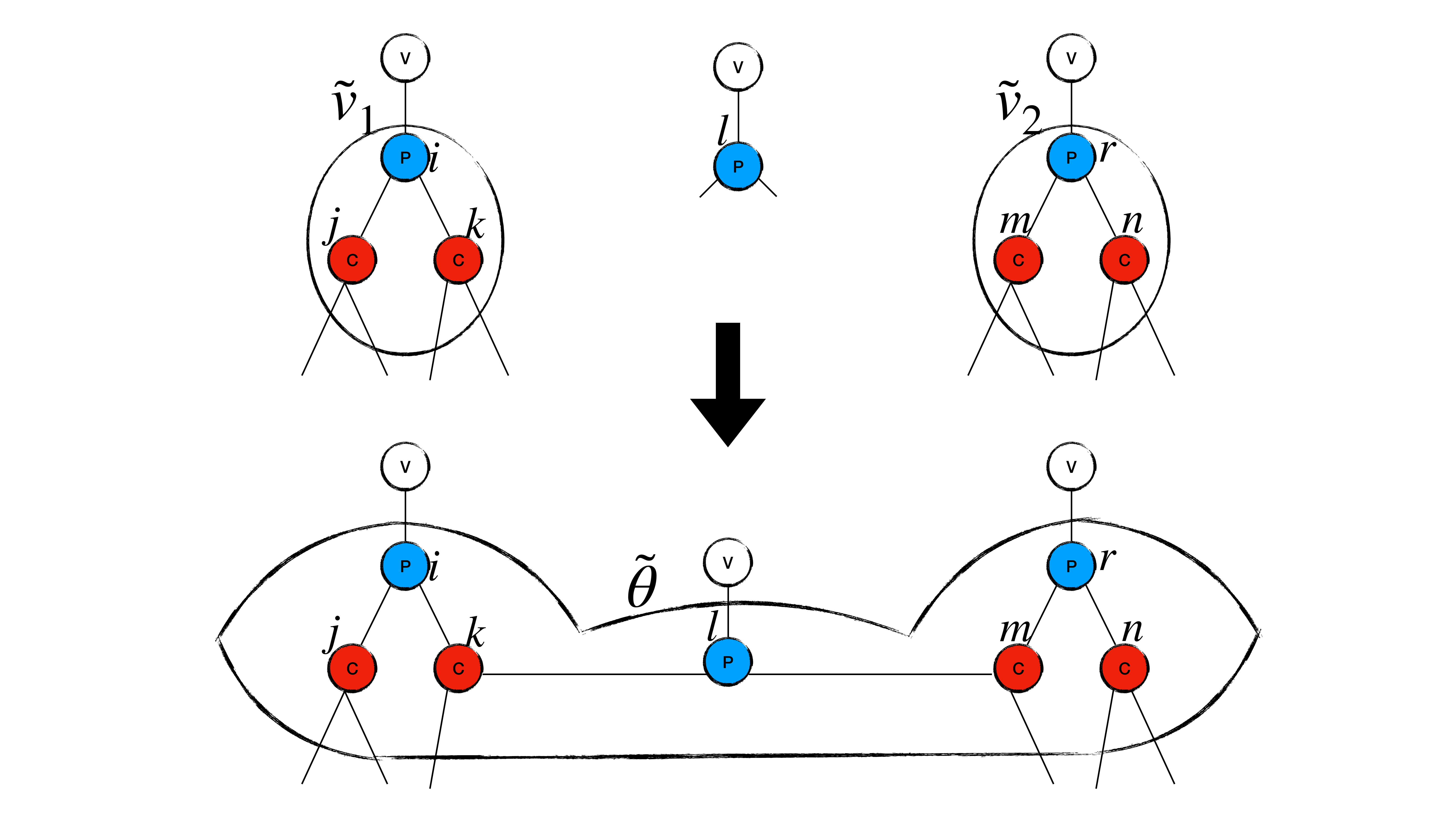}
    \captionsetup{font=footnotesize,justification=justified}
    \caption{\small The figure shows how a virtual site $\tilde{\theta}\in {\mathcal{A}}$ with $\overline{\Delta}_{\tilde{\theta}}=6$ is created. }
    \label{fig:chemic3}
\end{figure}

We define ${\mathcal{A}}$ to be  the set of virtual sites. The set  ${\mathcal{A}}$ is kept in ascending order respect to the $anti$-degree $\overline{\Delta}$ of each virtual site $\tilde{v}\in {\mathcal{A}}$. Virtual sites can be created, as described above, expanded or merged together (creating a new virtual site $\tilde{\theta}=\cup_i \tilde{v}_i $).  Two examples are shown in Fig. \ref{fig:chemic2} and \ref{fig:chemic3}.

Fig. \ref{fig:chemic2} shows how to expand a virtual site $\tilde{v}_1$. Let's imagine that a site $m \in {\mathcal{P}}$, with anti-degree $\overline{\Delta}_{m}=2$,  is chosen. Let's run Subroutine GA($m$, $\overline{\Delta}_m$) on $m$. Assume that $m$ connects with $n\in \mathcal{N}$ ($\overline{\Delta}_{n}=2$) and with $\tilde{v}_1\in {\mathcal{A}}$ ($\overline{\Delta}_{\tilde{v}_1}=4$). In this case $\tilde{v}_1\in {\mathcal{A}}$ expands itself, swallowing sites $m$ and $n$ and having a $\overline{\Delta}_{\tilde{v}_1}=5$.

 Fig. \ref{fig:chemic3} shows how two virtual sites merge together. Let's imagine that a site $p \in {\mathcal{P}}$, with anti-degree $\overline{\Delta}_{p}=2$,  is chosen during the building graph process.  Let's run Subroutine GA($p$, $\overline{\Delta}_p$) on $m$. Assume that $p$  connects with two virtual sites $\tilde{v}_1\in {\mathcal{A}}$ and $\tilde{v}_2\in {\mathcal{A}}$ with $\overline{\Delta}_{\tilde{v}_1}=4$ and $\overline{\Delta}_{\tilde{v}_2}=4$. The new structure is a virtual site $\tilde{\theta}\in {\mathcal{A}}$ with $\overline{\Delta}_{\tilde{\theta}}=6$.


\begin{figure}
    \centering
    \includegraphics[width=0.8\columnwidth]{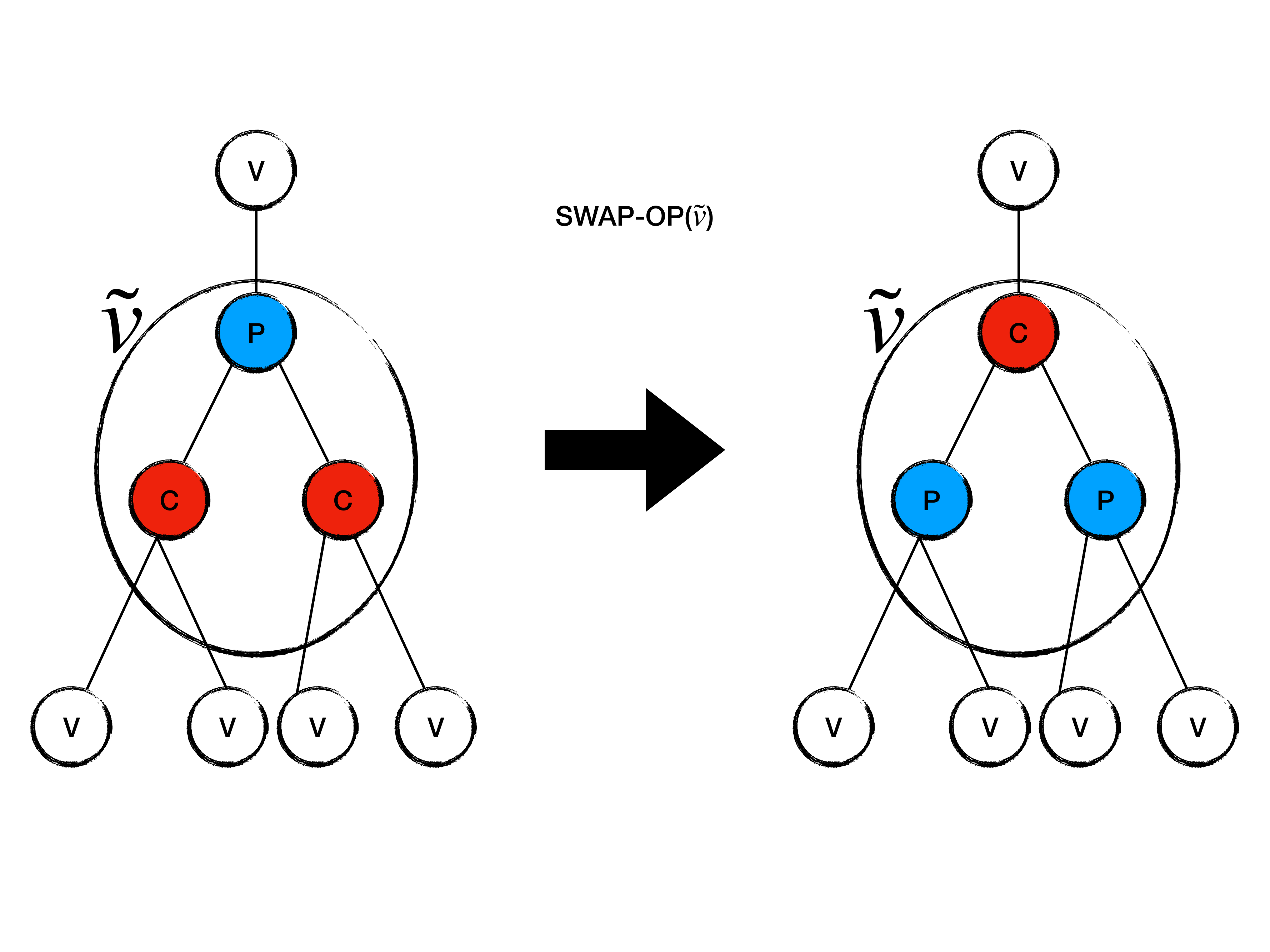}
    \captionsetup{font=footnotesize,justification=justified}
    \caption{\small The figure shows how $SWAP-OP(\tilde{v})$ works on a virtual site $\tilde{v}$, which has $\overline{\Delta}_{\tilde{v}}=0$. $SWAP-OP(\tilde{v})$ swaps all $P$ sites in $\tilde{v}$ into $C$ sites and all $C$ sites in $\tilde{v}$ into $P$ sites.} 
    \label{fig:swap}
\end{figure}

We define in the following a list of operations that will be useful for understanding the algorithm presented in the next sections.
\begin{definition}[$OP^i_{move}(i, \mathcal{X}, \mathcal{Y})$]
\textit{Let $\mathcal{X}$ and $\mathcal{Y}$ be two sets. Let $i\in \mathcal{X}$ and $i\notin \mathcal{Y}$. We define $OP^i_{move}(i, \mathcal{X}, \mathcal{Y})$ to be the operation that moves the site $i \in \mathcal{X}$ from the set $\mathcal{X}$ to $\mathcal{Y}$, i.e., $i\in \mathcal{Y}$ and $i \notin \mathcal{X}$. }
\end{definition}
For example, $OP^i_{move}(i, \mathcal{N}, \mathcal{V})$ moves $i\in \mathcal{N}$ from the set $\mathcal{N}$ to $\mathcal{V}$, i.e., $i\in \mathcal{V}$ and $i\notin \mathcal{N}$. Instead, the operation  $OP^i_{move}(i, \mathcal{N}, \mathcal{I})$ moves $i\in \mathcal{N}$ from the set $\mathcal{N}$ to $\mathcal{I}$, i.e., $i\in \mathcal{I}$ and $i\notin \mathcal{N}$. We recall that when a site is set into $\mathcal{I}$ it is labeled with $I$, while when a site is set into $\mathcal{V}$ it is labeled with $V$.

\begin{definition}[$OP^{\tilde{v}}_{del}(\tilde{v},  {\mathcal{A}})$]
\textit{Let ${\mathcal{A}}$ the set that contains virtual nodes $\tilde{v}$. We define $OP^{\tilde{v}}_{del}(\tilde{v}, {\mathcal{A}})$. to be the operation that deletes the site $\tilde{v} \in {\mathcal{A}}$ from the set ${\mathcal{A}}$,i.e., the element $\tilde{v} \notin {\mathcal{A}}$ anymore, and applies the operation $OP^i_{move}(i, \mathcal{X}, \mathcal{Y})$ on each site $i\in \tilde{v}$, following the rule: 
\begin{itemize}
\item if $i \in \tilde{v}$ is labeled with the letter $P$ then $\mathcal{X}=\mathcal{N}$ and $\mathcal{Y}=\mathcal{I}$;
\item if $i \in \tilde{v}$ is labeled with the letter $C$ then $\mathcal{X}=\mathcal{N}$ and $\mathcal{Y}=\mathcal{V}$.
\end{itemize}
}
\end{definition}

\begin{definition}[$SWAP-OP(\tilde{v})$]
\textit{Let $\tilde{v} \in {\mathcal{A}}$. We define $SWAP-OP(\tilde{v})$ to be the operation such that $\forall i \in \tilde{v}$:
\begin{itemize}
\item  if $i$ is labeled $P$ then the label $P$ swaps to $C$;
\item if $i$ is labeled $C$ then the label $C$ swaps to $P$;
\end{itemize}
Fig. \ref{fig:swap} shows how $SWAP-OP(\tilde{v})$ acts on a virtual site $\tilde{v}$.}
\end{definition}

\section{The deferred decision algorithm for $d=3$}
\label{sec:d=3}

In this section, we present our algorithm, simpler and slightly different from the one in \cite{csoka2016independent}, but based on the same idea, for determining large independent set in random $d$-regular graphs with $d=3$, i.e., $G_3(N)$.
It will also be the core of the algorithm developed in Sec. \ref{sec:d>3}. As mentioned above, the algorithm discussed in this paper is basically a prioritized algorithm, i.e., algorithms that make local choices in which there is a priority in selecting a certain site. Our algorithm belongs to this class.

We start the discussion on the local algorithm for $d=3$ by giving the pseudo-code of the algorithm in \textbf{ Algorithm~\ref{algoL3}}. 

The algorithm starts by randomly taking a site $i$ from the set $\mathcal{N}$ and completes its connections in a random way, following the method described in \textbf{Algorithm \ref{algo1}}. 
Once all its connections are completed, site $i$ has $\Delta_i=3$ and $\overline{\Delta}_i = 0$. It is labeled with letter $V$, erased from $\mathcal{N}$, and set into $\mathcal{V}$. In other words, operation $OP^i_{move}(i, \mathcal{N}, \mathcal{V})$ is applied on it. Each neighbor of $i$, i.e., $j\in \partial i$, has degree $\Delta_j=1$ and anti-degree $\overline{\Delta}_j=2$. Therefore, they are set into ${\mathcal{P}}$, thus labeled $P$.

The algorithm picks a site $k$ from ${\mathcal{P}}$ with the minimum remaining connections. 
In general, If $k$ has $\overline{\Delta}_k\neq0$,  the algorithm completes all its connections, and removes it from ${\mathcal{P}}$. Each site connected with a $P$ is automatically labeled with the letter $C$. If a site $k \in {\mathcal{P}}$ connects to another site $j \in {\mathcal{P}}$, with $j \neq k$, $j$ is removed from ${\mathcal{P}}$ and it is labeled $C$.

If $ k \in {\mathcal{P}}$ has $\overline{\Delta}_k=0$, the site $k$ is set into $\mathcal{I}$, and it is removed from $\mathcal{N}$ and ${\mathcal{P}}$, i.e. the algorithm applies the operation $OP^k_{move}(k, \mathcal{N}, \mathcal{I})$.

As defined in Sec.~\ref{sec:definition}, a $C-P-C$ structure is equivalent to a single \textit{virtual} site, $\tilde{v}$, which has an anti-degree $\overline{\Delta}_{\tilde{v}}$. Each virtual site $\tilde{v}$ created with $\overline{\Delta}_{\tilde{v}}>2 $, is inserted into the set ${\mathcal{A}}$.

Once the set ${\mathcal{P}}$ is empty, the algorithm selects a site $\tilde{v}\in {\mathcal{A}}$ with the largest anti-degree $\overline{\Delta}_{\tilde{v}}$, and it applies the operation $OP^{\tilde{v}}_{del}(\tilde{v}, {\mathcal{A}})$ after having completed all the connections  $\forall i \in \tilde{v}$ with $\overline{\Delta}_{i}\neq 0$, using on each $i  \in \tilde{v}$ with $\overline{\Delta}_{i}\neq 0$ \textbf{Algorithm} \ref{algo1}. 

We apply operation $OP^{\tilde{v}}_{del}(\tilde{v}, {\mathcal{A}})$ on virtual sites $\tilde{v} \in \mathcal{A}$ with the largest anti-degree because we hope the random connections outgoing from those sites could reduce the  anti-degrees of existing virtual sites in $\mathcal{A}$, in such a way that the probability to have  virtual nodes with anti-degree $\overline{\Delta}_{\tilde{v}} \leq 2$ increases. In other words, we want to create islands of virtual sites that are surrounded by a sea of $V$ sites for applying the $SWAP-OP(\tilde{v})$ on those nodes. This protocol, indeed, allows to increase the independent set cardinality and decrease the vertex cover set cardinality. 


For this reason, if virtual nodes with anti-degree $\overline{\Delta}_{\tilde{v}} \leq 2$ exist in ${\mathcal{A}}$, those sites have the highest priority in being selected. More precisely, the algorithm follows the priority rule:
\begin{enumerate}

\item $\forall \tilde{v} \in {\mathcal{A}}$ s.t. $\overline{\Delta}_{\tilde{v}}=0$ the algorithm applies sequentially the operation $SWAP-OP(\tilde{v})$ and then the operation $OP^{\tilde{v}}_{del}(\tilde{v}, {\mathcal{A}})$.

\item If no virtual sites $\tilde{v} \in {\mathcal{A}}$ with $\overline{\Delta}_{\tilde{v}} = 0$ are present, then the algorithm looks for those that have $\overline{\Delta}_{\tilde{v}} = 1$. $\forall \tilde{v} \in {\mathcal{A}}$ s.t. $\overline{\Delta}_{\tilde{v}}=1$ it applies the operation $SWAP-OP(\tilde{v})$, completes the last connection of the site $i\in \tilde{v}$ with $\overline{\Delta}_{i}=1$, applies on the last neighbour of $i$ added $OP^j_{move}(j, \mathcal{N}, \mathcal{V})$, and then $OP^{\tilde{v}}_{del}(\tilde{v}, {\mathcal{A}})$.

\item If no virtual sites $\tilde{v} \in {\mathcal{A}}$ with $\overline{\Delta}_{\tilde{v}} = 0$ and $\overline{\Delta}_{\tilde{v}} = 1$ are present, then the algorithm looks for those that have $\overline{\Delta}_{\tilde{v}} = 2$.  $\forall \tilde{v} \in {\mathcal{A}}$ s.t. $\overline{\Delta}_{\tilde{v}}=2$ it applies the operation $SWAP-OP(\tilde{v})$, completes the last connections of the sites $i\in \tilde{v}$ with $\overline{\Delta}_{i}\neq 0$, labels the new added sites with the letter $C$, and updates the degree and the anti-degree of the virtual node $\tilde{v}$.

\end{enumerate}

\begin{figure}
  \centering
  \includegraphics[width=0.8\columnwidth]{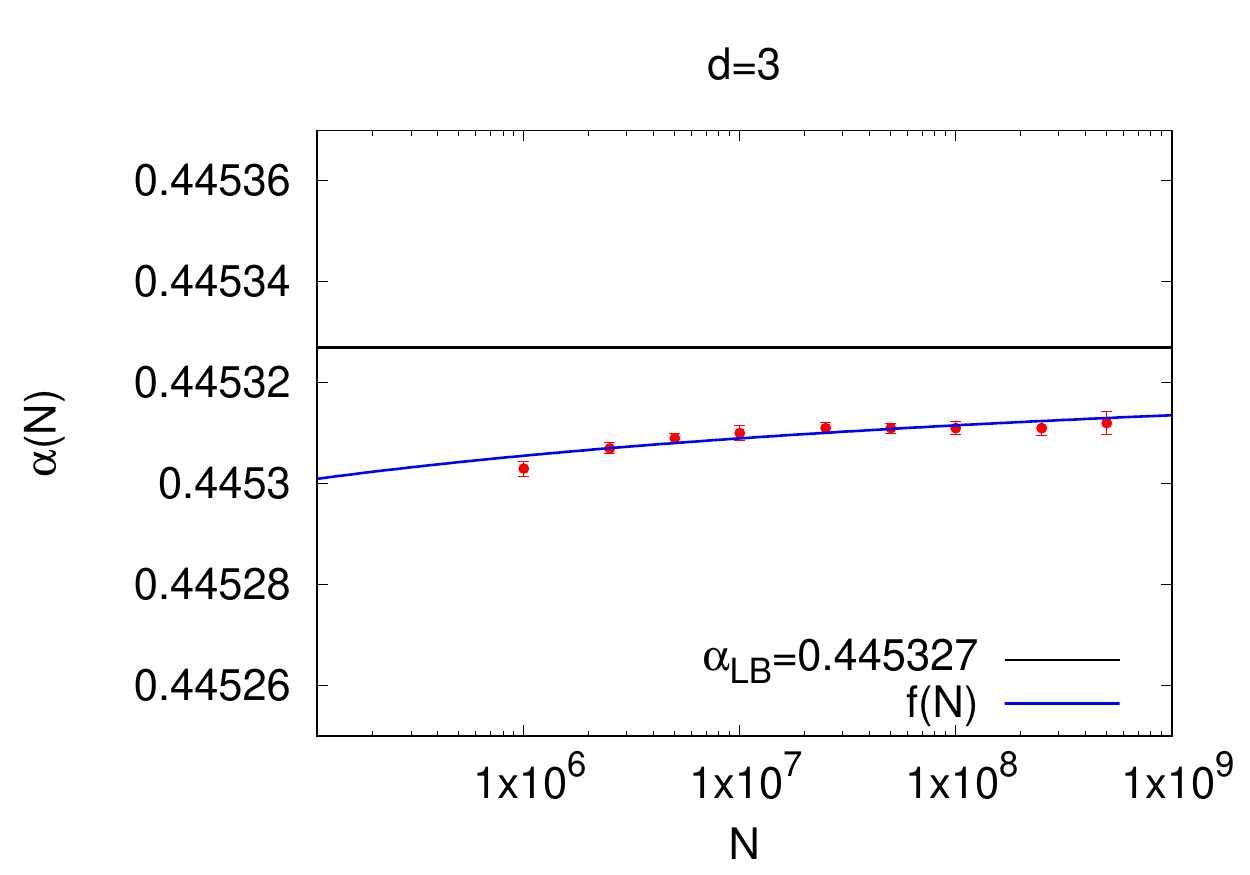}
  \captionsetup{font=footnotesize,justification=justified}
  \caption{\small The figure shows the extrapolation of the independent set ratio as a function of graphs order, i.e. $|\mathcal{N}|=N$, and $d=3$. The error bars identify the standard errors multiplied for the quantile of the t distribution $z_{99\%}=3.35$. 
  The asymptotic value $\alpha_{\infty}= 0.445330 (3)$, extrapolated by fitting the data using a function $f(N)=(a/ \ln N)+\alpha_{\infty}$ (blue line), where $a=-0.00033(6)$, identifies the value that our algorithm can reach when $n \to \infty$. The confidence interval at $99\%$ of $\alpha_{\infty}$ (in Tab. $1$) agrees with the theoretical value of $\alpha_{LB}$=0.44533. }
  \label{fig:d3vsN}
\end{figure}

The algorithm proceeds selecting virtual nodes and creating sites labeled $P$ until $\mathcal{N}=\emptyset$. Once  $\mathcal{N}=\emptyset$ it returns the set $\mathcal{I}$. The set of independent sites.
The code of the algorithm can be downloaded at \cite{GitHubCode}. 

\begin{table}
\label{tab-d=3}
\centering
\begin{tabular}{ |c||c|c|} 
\hline
 $N$ & $d$ & $\alpha (N) \pm \sigma_{\alpha(N)}$\\
\hline
$10^6$ &3& 0.445303 (48)\\ 
$2.5 \cdot 10^6$ &3& 0.445307 (30)\\ 
 $5 \cdot 10^6$ &3& 0.445309 (21)\\ 
$10^7$ &3& 0.445310 (15) \\
$2.5 \cdot 10^7$ &3& 0.445311 (9) \\ 
$5 \cdot 10^7$ &3& 0.445311 (7) \\ 
$10^8$ &3& 0.445311 (4) \\
$2.5 \cdot 10^8$ &3& 0.445311 (4)\\ 
$5 \cdot 10^8$ &3& 0.445312 (2)\\ 
\hline
\end{tabular}
\captionsetup{font=footnotesize,justification=justified}
\caption{The table shows the sample average and standard deviation values of the independent set ratio $\alpha(N)$ the for random regular graphs of order $N$ and $d=3$.}
\end{table}

We are comparing numerical results for independence ratios that agree with theoretical ones, at least, up to $5^{\text{th}}$ digit. For this reason, we performed an accurate analysis on random $3$-regular graphs starting from those that have an order of $10^6$, and pushing it up to $5\cdot10^8$. 
This analysis aims to compute the sample mean of the independence ratio size $\alpha(N)$ outputted by our algorithm. Each average is obtained in the following manner: for graphs of order $N = 10^6$ we averaged over a sample of $10^4$ graphs; for order $N = 2.5 \cdot 10^6$ we make an average over a sample of $7.5\cdot 10^3$ graphs; for order $N = 5\cdot 10^6$ we make an average over a sample of $5\cdot 10^3$ graphs; for order $N = 10^7$ the average is performed over a sample of $10^3$ graphs; for $N = 2.5\cdot 10^7$ over $7.5\cdot 10^2$ graphs, for $N = 5\cdot 10^7$ over $5\cdot 10^2$ graphs, for $N = 10^8$ over $10^2$ graphs, for $N = 2.5\cdot 10^8$ over $50$ graphs, and for $N = 5\cdot 10^8$ over $10$ graphs. The mean and the standard deviation for each sample analyzed are reported in Tab. $2$. Observing that the values of each independent set ratio sample mean reach an asymptotic value, we perform a linear regression on the model $f(N)=(a/ \ln N)+\alpha_{\infty}$ for estimating the parameter $\alpha_{\infty}$ (blue line in Fig. \ref{fig:d3vsN}). When $N \to \infty$ the first term of the regression, i.e., $(a/ \ln N)$, goes to $0$ leaving out the value of $\alpha_{\infty}$ that describes the asymptotic value of the independence ratio that our algorithm can reach. Using the numerical standard errors obtained on each sample, we apply a General Least Square (GLS) method \cite{bishop2006pattern} for inferring the values of the parameters $\alpha_{\infty}$,  averaging sufficient simulations to achieve a confidence  interval at $99\%$ on it. The value of $\alpha_{\infty}$ is the most important because it is the asymptotic value that our algorithm can reach when $N \to \infty$. From the theory of GLS, we know that the estimator of parameter $\alpha_{\infty}$ is unbiased, consistent and efficient, and a confidence interval on this parameter is justified. The analysis, performed on data reported in Tab. $2$, shows that the independent set ratio reaches the asymptotic value $\alpha_{\infty}= 0.445330 (3)$. This value agrees with the theoretical value proposed in \cite{csoka2016independent}.

\begin{algorithm}[H]
 \label{algoL3}

\KwInput{$N$, $d=3$;}
\KwOutput{$\mathcal{I}$;}
Build the set of sites $\mathcal{N}$ with  $|\mathcal{N}|=N$;\\
$\mathcal{I}=\emptyset$;\\
$\mathcal{V}=\emptyset$;\\
Pick a random site $i \in \mathcal{N}$;\\
Run Subroutine $GA(i, \overline{\Delta}_i)$;\\
Apply $OP^i_{move}(i, \mathcal{N}, \mathcal{V})$;\\
\While{$\mathcal{N}\neq \emptyset$}{
\While{ $\exists i \in \mathcal{N}$ s.t. $\overline{\Delta}_i\leq2$ $\land$  $i \notin {\mathcal{P}}$ $\land$ $i$ is not labeled $C$}{
	 Label $i$ with letter $P$ and insert it into  ${\mathcal{P}}$;\\
	 }
\If{${\mathcal{P}}\neq \emptyset$}{
	\While{${\mathcal{P}} \neq \emptyset$}{
	Pick the first $l\in {\mathcal{P}}$ (we recall that elements in ${\mathcal{P}}$ are in ascending order);\\
	\If{$\overline{\Delta}_{l}=0$}{
	Apply $OP^l_{move}(l, \mathcal{N}, \mathcal{I})$;\\
	Remove $l$ from ${\mathcal{P}}$;\\
	}\Else{
	Run Subroutine  $GA(l, \overline{\Delta}_l)$;\\
	If a neighbour $j$ of $l$, i.e., $j \in \partial l$, is in ${\mathcal{P}}$ remove $j$ from ${\mathcal{P}}$;\\
	$\forall j \in \partial l$, label each $j$ with the letter $C$;\\
	Build or update the $virtual$ node $\tilde{v}$ and, if it is not present, insert it into ${\mathcal{A}}$;\\
	Remove $l$ from ${\mathcal{P}}$;\\
	}
	}
}\Else{
	 \While{$\exists \tilde{v}\in {\mathcal{A}}$ s.t. $\overline{\Delta}_{\tilde{v}}=0$}{
	 Apply $SWAP-OP(\tilde{v})$;\\
	 Apply $OP^{\tilde{v}}_{del}(\tilde{v},  {\mathcal{A}})$;\\
	 } 
	 \While{$\exists \tilde{v}\in {\mathcal{A}}$ s.t. $\overline{\Delta}_{\tilde{v}}=1$}{
	 Apply $SWAP-OP(\tilde{v})$;\\
	 For $i\in \tilde{v}$ labeled $P$ s.t. $\overline{\Delta}_{i}=1$ run Subroutine $GA(i, \overline{\Delta}_i)$;\\
	 Pick $j \in \partial i$,with $j$ the last neighbour of $i$ added;\\
	Run Subroutine  $GA(j, \overline{\Delta}_j)$;\\
	 Apply $OP^j_{move}(j, \mathcal{N}, \mathcal{V})$;\\
	 Apply $OP^{\tilde{v}}_{del}(\tilde{v},  {\mathcal{A}})$;\\
	 }
	  \While{$\exists \tilde{v}\in {\mathcal{A}}$ s.t. $\overline{\Delta}_{\tilde{v}}=2$}{
	 Apply $SWAP-OP(\tilde{v})$;\\
	 $\forall i\in \tilde{v}$ labeled $P$ s.t. $\overline{\Delta}_{i} \leq 2$ run Subroutine $GA(i, \overline{\Delta}_i)$ and label the neighbour(s) of $i$ with the letter $C$;\\
	 Update the $virtual$ node $\tilde{v}$;\\
	 }
	Pick $\tilde{v}$ s.t. $\max_{\tilde{v}\in {\mathcal{A}}} \overline{\Delta}_{\tilde{v}}$;\\
	$\forall i\in \tilde{v}$ s.t. $\overline{\Delta}_{i} \neq 0$ and labeled $C$,  run Subroutine $GA(i, \overline{\Delta}_i)$;\\
	Apply $OP^{\tilde{v}}_{del}(\tilde{v},  {\mathcal{A}})$;\\ 
}
}
 \Return $\mathcal{I}$;
 \caption{local algorithm for $d=3$}
\end{algorithm}

\section{The deferred decision algorithm for $d>3$}
\label{sec:d>3}

In this section, we present how to generalize the  prioritized algorithm for all $d>3$. It, like the one previously described in Sec. \ref{sec:d=3}, builds the random regular graph, and, at the same time, tries to maximize the independent set cardinality $|\mathcal{I}|$. 
The main idea that we propose is to melt down two existing algorithms, namely the one in \cite{wormald1995differential} and the one above described, into a new prioritized algorithm, which  is able to maximize the independent set cardinality, providing improved estimates of lower bounds.    The new conjectured lower bounds  come from extrapolation on random  $d$-regular graphs of size up to $10^9$. 

Before introducing the algorithm, we present a new operation that will allow us to simplify the discussion.

\begin{definition}[$OP^{i}_{build-del}(i, \mathcal{N}, \mathcal{I}, \mathcal{V})$]
\textit{Let  $i \in \mathcal{N}$. We define $OP^{i}_{build-del}(i, \mathcal{N}, \mathcal{I}, \mathcal{V})$ the operation that connects $i$ to $\overline{\Delta}_i$ sites following \textbf{Algorithm~\ref{GA}} rules, applies $OP^i_{move}(i, \mathcal{N}, \mathcal{I})$, and $\forall j\in \partial i$ sequentially runs \textbf{Algorithm~\ref{GA}} and applies the operation $OP^j_{move}(j, \mathcal{N}, \mathcal{V})$.} 
\end{definition}

The pseudo-code of the last operation is described in \textbf{Algorithm~\ref{op5}}.

\begin{algorithm}[H]
 \label{op5}
\KwInput{$i\in \mathcal{N}$, $\mathcal{N}$, $\mathcal{I}$,  $\mathcal{V}$;}
\KwOutput{$\mathcal{N}$, $\mathcal{I}$, $\mathcal{V}$;}
 
Run Subroutine $GA(i, \overline{\Delta}_i)$;\\
Apply $OP^i_{move}(i, \mathcal{N}, \mathcal{I})$;\\
 \While{$\partial i \neq \emptyset$}{
Pick $j\in \partial i$;\\
Run Subroutine $GA(j, \overline{\Delta}_j)$;\\
Apply $OP^j_{move}(j, \mathcal{N}, \mathcal{V})$;\\
 }
  \Return $\mathcal{N}$, $\mathcal{I}$, $\mathcal{V}$;
 \caption{$OP^{i}_{build-del}(i, \mathcal{N}, \mathcal{I}, \mathcal{V})$}
 \end{algorithm}

We start the discussion on the local algorithm for $d>3$ by giving the pseudo-code of the algorithm in \textbf{ Algorithm~\ref{algoLb3}}. 

\begin{figure}
    \centering
    \includegraphics[width=0.8\columnwidth]{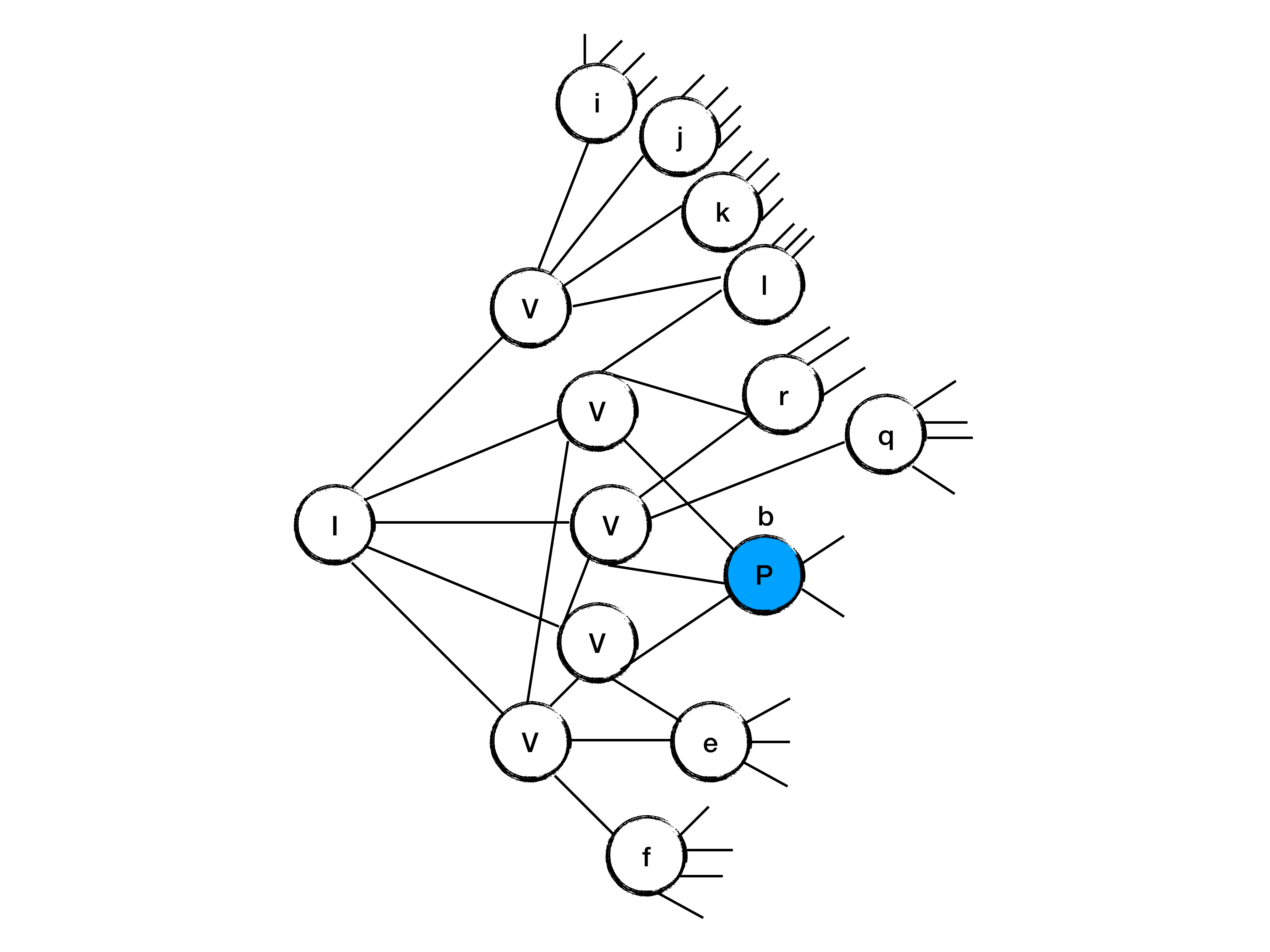}
    \captionsetup{font=footnotesize,justification=justified}
    \caption{\small The figure shows how a $P$ site appears, i.e. a site $b$ with $\overline{\Delta}_{b}\leq2$, in a random $d$-regular graph $ G_d(\mathcal{N},E)$ with degree $d=5$. In the event that a $P$ site has not been created, the algorithm picks a site $ m \in \mathcal{N}$ with minimum $\overline{\Delta}_{m}$ and applies operation  $OP^{m}_{build-del}(m, \mathcal{N}, \mathcal{I}, \mathcal{V})$ on it.} 
    \label{fig:p-d5}
\end{figure}

\begin{figure}
\label{fig:dynamic_p}
    \centering
    \includegraphics[width=0.7\columnwidth]{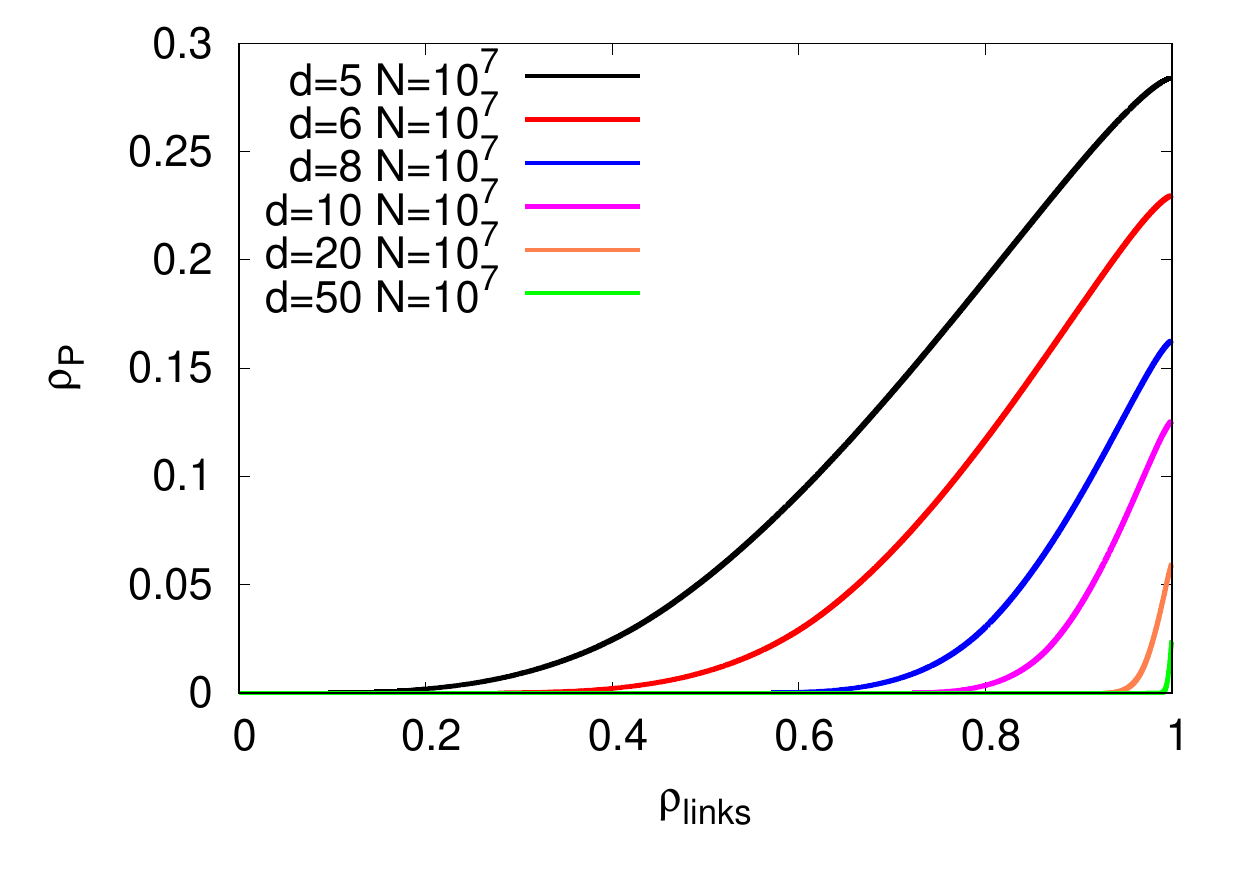}
     \includegraphics[width=0.7\columnwidth]{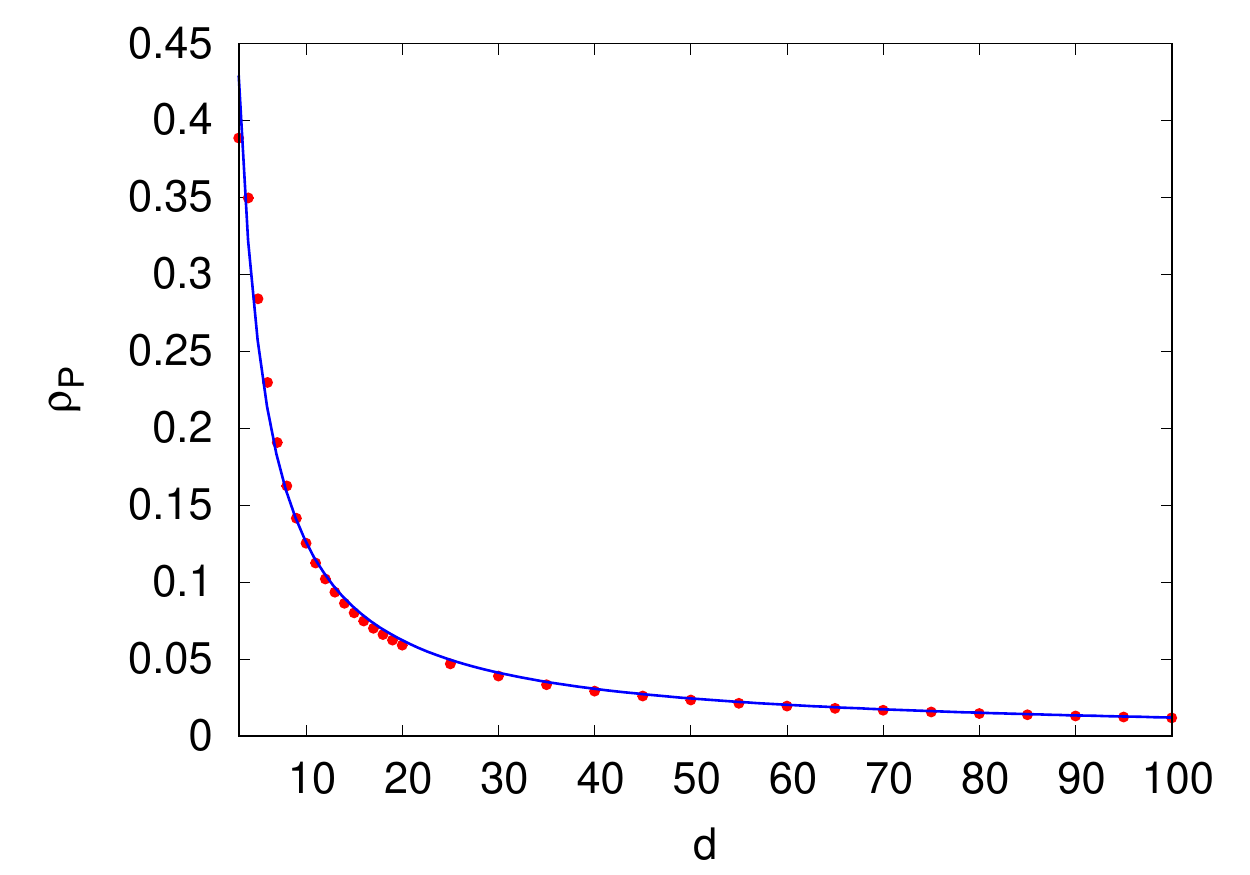}
    \captionsetup{font=footnotesize,justification=justified}
    \caption{\small The figure shows how the dynamics of the fraction of $P$ sites in the graph building process appears as a function of links inserted in (top panel), and the total fraction of $P$ sites as a function of $d$ (lower panel). The fraction of $P$ sites decreases like $\sim 1/d$, when $d$ becomes large. As stated in the main text this behavior shows that for $d \to \infty$ our algorithm matches the one in \cite{wormald1995differential}.} 
\end{figure}

The algorithm starts randomly selecting a site $z$  from the set of all nodes $\mathcal{N}$, i.e. $z \in \mathcal{N}$.
It then applies $OP^{z}_{build-del}(z, \mathcal{N}, \mathcal{I}, \mathcal{V})$ on the site $z$ (see Fig. \ref{fig:p-d5}). 
This operation creates nodes with different degrees and anti-degrees. 
The algorithm proceeds in choosing the node $m$, from those with minimum $\overline{\Delta}_m$. 
If the node $m$ has $\overline{\Delta}_m> 2$, the algorithm applies the operation $OP^{m}_{build-del}(m, \mathcal{N}, \mathcal{I}, \mathcal{V})$ on site $m$.  In other words, we are using the algorithm developed in \cite{wormald1995differential} until a site $m \in \mathcal{N}$ with $\overline{\Delta}_m\leq 2$ pops up. When such a case appears, we label it as a $P$ site and we move it into the set ${\mathcal{P}}$.

As before, once ${\mathcal{P}}$ is not empty, the sites in ${\mathcal{P}}$ have the highest priority in being processed for creating virtual nodes. 

In principle more complex virtual nodes can be created. For instance defining a $P$ site as a site $i \in \mathcal{N} $ not labeled yet with any letter ($C$ or $P$) s.t.  $\overline{\Delta}_j\leq d-1$, with ${\Delta}_j$ random connections with sites already in $\mathcal{V}$.  Although we do not see any logical impediment in creating more complex virtual nodes, we confine ourselves to the case where the anti-degree of a bare site in $\mathcal{N}$ is less or equal to two for any values of $d$, because it is much easier to handle and explain.

Until the set ${\mathcal{P}}$ is empty, the algorithm builds virtual sites, which are set into ${\mathcal{A}}$. 
 
Once the set ${\mathcal{P}}$ is empty, the highest priority in being covered is placed on the virtual sites contained in ${\mathcal{A}}$.
The rules that the algorithm follows are:

\begin{enumerate}

\item $\forall \tilde{v} \in {\mathcal{A}} \text{  s.t.   } \overline{\Delta}_{\tilde{v}}=0 \lor \overline{\Delta}_{\tilde{v}}=1$, the algorithm applies sequentially operation $SWAP-OP(\tilde{v})$  and the operation $OP^{\tilde{v}}_{del}(\tilde{v}, {\mathcal{A}})$. (in the case $\overline{\Delta}_{\tilde{v}}=1$, the algorithm before it completes the absent  connection for the site $i \in \tilde{v}$ s.t. $\overline{\Delta}_i=1$, and it applies on the last added site $j\in \partial i$ the operation $OP^j_{move}(j, \mathcal{N}, \mathcal{V})$. Then on the virtual site $\tilde{v}$ it applies $OP^{\tilde{v}}_{del}(\tilde{v},  {\mathcal{A}})$).

\item If $\exists \tilde{v} \in {\mathcal{A}} \text{ s.t. } \overline{\Delta}_{\tilde{v}}= 2$ $\land$ $\nexists \tilde{q} \in {\mathcal{A}} \text{ s.t. } \overline{\Delta}_{\tilde{q}} = 0 \lor \overline{\Delta}_{\tilde{q}}=1$ the algorithm chooses with the highest priority the site with $\overline{\Delta}_{\tilde{v}}=2$. Then it applies operation $SWAP-OP(\tilde{v})$ on $\tilde{v}$, it runs $\forall i\in \tilde{v}$ with $\overline{\Delta}_i\neq 0$ the Subroutine $GA(i, \overline{\Delta}_i)$ and labels each neighbour(s) of $i$ with letter $C$.

\item If $\exists \tilde{v} \in {\mathcal{A}} \text{ s.t. } \overline{\Delta}_{\tilde{v}}> 2$  $\land$ $\nexists \tilde{p} \in {\mathcal{A}} \text{ s.t. } \overline{\Delta}_{\tilde{p}}\leq 2$ the algorithm chooses a  a site $\tilde{v} \in {\mathcal{A}}$ with the maximum $\overline{\Delta}_{\tilde{v}}$ and it applies the $OP^{\tilde{v}}_{del}(\tilde{v},  {\mathcal{A}}) $ with the maximum $\overline{\Delta}_{\tilde{v}}$, after having run the Subroutine $GA(i, \overline{\Delta}_i)$ on each $i\in \tilde{v}$ such that $\overline{\Delta}_i\neq 0$.

\item In the case ${\mathcal{P}}=\emptyset \land {\mathcal{A}}=\emptyset \land \mathcal{N}\neq \emptyset$  the algorithm takes a site $t \in \mathcal{N}$ with minimum $ \overline{\Delta}_t $, and applies the operation $OP^{t}_{build-del}(t, \mathcal{N}, \mathcal{I}, \mathcal{V})$.

\end{enumerate}

The algorithm works until the following condition is true: $\mathcal{N}= \emptyset \land {\mathcal{P}}=\emptyset \land  {\mathcal{A}}=\emptyset$. Then, it checks that all sites in $\mathcal{I}$ are covered only by sites in $\mathcal{V}$, and no site in $\mathcal{I}$ connects to any other site in $\mathcal{I}$. 

The results obtained by the algorithm at different values of $d$, and at different orders $N$, are presented in Tab. $3$, $4$, and $5$. The confidence intervals of the asymptotic  independent set ratio values, obtained by extrapolation described in the previous section, are presented in Tab. $1$. In other words, we performed simulations for each value of $d$ by computing the sample mean and the standard error of the independence ratio a some values of $N$. Then we use GLS methods for extrapolating  the values of $\alpha_{\infty}$ and building up its confidence interval.

\begin{table}
\label{tab-d=4-5-6-7}
\centering
\begin{tabular}{ |c||c|c||c|c||c|c||c|c| } 
\hline
 $N$ &  $d$ &  $\alpha (N) \pm \sigma_{\alpha(N)}$ & $d$ &  $\alpha (N) \pm \sigma_{\alpha(N)}$ & $d$ &  $\alpha (N) \pm \sigma_{\alpha(N)}$ & $d$ &  $\alpha (N) \pm \sigma_{\alpha(N)}$\\
\hline
$10^6$ &4  & 0.400831 (66) &5 &  0.364723 (78) &6 & 0.335964 (84) &7 & 0.312367 (89)\\ 
$2.5\cdot 10^6$ &4 &  0.400837 (41) &5& 0.364731 (48)&6 & 0.335969 (53) &7 & 0.312373 (56)\\ 
 $5 \cdot 10^6$ &4 &  0.400840 (30) &5& 0.364732 (35) &6 &  0.335972 (38) &7 & 0.312378 (37)\\ 
$10^7$ &4 & 0.400840 (21) &5& 0.364733 (24)&6 &  0.335975 (26) &7 &  0.312378 (27)\\
$2.5\cdot 10^7$ &4 & 0.400841 (13) &5& 0.364734 (15) &6 &   0.335976 (18) &7 & 0.312380 (18)\\ 
$5 \cdot 10^7$ &4 & 0.400843 (9) &5& 0.364735  (11)&6 &   0.335975 (12) &7 & 0.312381 (13)\\ 
$10^8$ &4&  0.400842 (6) &5& 0.364734 (7)&6 &   0.335975 (8) &7 & 0.312381 (9)\\
$2.5\cdot 10^8$ &4 & 0.400843 (4) &5  & 0.364735 (5)&6 &   0.335975 (6) &7 & 0.312381 (5)\\ 
$5 \cdot 10^8$ &4 & 0.400842 (3) &5  & 0.364734 (3) &6 &   0.335977 (2) &7 & 0.312381 (4)\\ 
\hline
\end{tabular}

\captionsetup{font=footnotesize,justification=justified}
\caption{The table shows the sample average and standard deviation values of the independent set ratio $\alpha(N)$ for random regular graphs of order $N$ and degree  $d=4,\,5\,, 6\,, 7$.}
\end{table}
\begin{table}
\label{tab-d=8-9-10}
\centering

\begin{tabular}{ |c||c|c||c|c||c|c| } 
\hline
 $N$ & $d$ &  $\alpha (N) \pm \sigma_{\alpha(N)}$ & $d$ &  $\alpha (N) \pm \sigma_{\alpha(N)}$ & $d$ &  $\alpha (N) \pm \sigma_{\alpha(N)}$ \\
\hline
$10^6$ &8& 0.292522 (83) &9  & 0.275511 (85) &10 & 0.260747 (84)\\ 
$2.5\cdot 10^6$ &8& 0.292523 (53)&9 &  0.275517 (53) &10&  0.260753 (54)\\ 
 $5 \cdot 10^6$ &8& 0.292526 (37) &9 &  0.275519 (38) &10&  0.260755 (38)\\ 
$10^7$ &8& 0.292527 (26) &9 &0.275521 (27) &10&  0.260757 (28)\\
$2.5\cdot 10^7$ &8& 0.292529 (17) &9 & 0.275522 (17) &10& 0.260759 (17) \\ 
$5 \cdot 10^7$ &8& 0.292530 (11) &9 & 0.275521 (12) &10& 0.260758 (11)  \\ 
$10^8$ &8& 0.292530 (8) &9&  0.275523 (8) &10& 0.260759 (8)\\
$2.5\cdot 10^8$ &8& 0.292530 (4) &9 & 0.275523 (5) &10  &  0.260759 (5) \\ 
\hline
\end{tabular}

\captionsetup{font=footnotesize,justification=justified}
\caption{The table shows the sample average and standard deviation values of the independent set ratio  $\alpha(N)$ for random regular graphs of order $N$ and degree  $d=8,\,9,\, 10\,$. }
\end{table}
\begin{table}
\label{tab-d=20-50-100}
\centering

\begin{tabular}{ |c||c|c||c|c||c|c| } 
\hline
 $N$ & $d$ &  $\alpha (N) \pm \sigma_{\alpha(N)}$ & $d$ &  $\alpha (N) \pm \sigma_{\alpha(N)}$ & $d$ &  $\alpha (N) \pm \sigma_{\alpha(N)}$ \\
\hline
$2.5\cdot 10^5$ &20&  0.175389 (151) &50  & 0.095673 (114) &100 & 0.057523 (88) \\
$5\cdot 10^5$ &20&  0.175403 (107) &50  & 0.095682 (81) &100 &  0.057522 (62)\\
$10^6$ &20&  0.175407 (75) &50  &0.095684 (57) &100 & 0.057524 (43) \\ 
$2.5\cdot 10^6$ &20& 0.175412 (48) &50 & 0.095688 (36) &100& 0.057525 (27)\\ 
$5 \cdot 10^6$ &20& 0.175414 (33) &50 & 0.095689 (24)  &100& 0.057527 (20)\\ 
$10^7$ &20& 0.175415 (24) &50 & 0.095690 (18) &100& 0.057528 (14) \\
$2.5\cdot 10^7$ &20& 0.175416 (17)  &50 & 0.095691 (12) &100& 0.057527 (9) \\ 
$5 \cdot 10^7$ &20& 0.175418 (11) &50 & 0.095690 (9) &100& 0.057527 (7) \\ 
\hline
\end{tabular}

\captionsetup{font=footnotesize,justification=justified}
\caption{The table shows the sample average and standard deviation values of the independent set ratio $\alpha(N)$ for random regular graphs of order $N$ and degree  $d=20,\,50,\,100$. }
\end{table}

From our analysis, we observe that $\forall d>4$ our results, as far as we know, exceed the best theoretical lower bounds given by greedy algorithms. Those improvements are obtained because we 
allow the virtual nodes to increase and decrease their anti-degree. In other words, this process transforms the random  $d$-regular graph into a sparse random graph, where it is much easier making local rearrangements (our $SWAP-OP(\cdot)$ move)  to enlarge the independent set.

More precisely, the creation of virtual nodes that increase or decrease their anti-degrees allows us to deal with a graph that is not anymore $d$-regular but has average connectivity $\langle d \rangle$. 

However, this improvement decreases as $d$ becomes large, $\sim 1/d$, and disappears when $d \to \infty$ (see Fig. $7$, bottom panel). Indeed, the number of $P$ labeled sites decreases during the graph building process (see Fig. $7$, top panel), invalidating the creation of virtual nodes that are the core of our algorithm. This means that our algorithm for $d \to \infty$ will reach the same asymptotic independent set ratio values obtained by the algorithm in \cite{wormald1995differential}.

In conclusion, for any fixed and small $d$, we have that the two algorithms are distinct, and our algorithm produces better results without increasing the computational complexity.

\begin{algorithm}[H]
 \label{algoLb3}

\KwInput{$N$, $d$;}
\KwOutput{$\mathcal{I}$;}
Build the set of sites $\mathcal{N}$ with  $|\mathcal{N}|=N$;\\
$\mathcal{I}=\emptyset$;\\
$\mathcal{V}=\emptyset$;\\
Pick a random site $i \in \mathcal{N}$;\\
Apply $OP^{i}_{build-del}(i, \mathcal{N}, \mathcal{I}, \mathcal{V})$;\\
\While{$\mathcal{N}\neq \emptyset$}{
\While{ $\exists i \in \mathcal{N}$ s.t. $\overline{\Delta}_i\leq2$ $\land$  $i \notin {\mathcal{P}}$ $\land$ $i$ is not labeled $C$}{
	 Label $i$ with letter $P$ and insert it into  ${\mathcal{P}}$;\\
	 }
\uIf{${\mathcal{P}}\neq \emptyset$}{
	\While{${\mathcal{P}} \neq \emptyset$}{
	Pick the first $l\in {\mathcal{P}}$;\\
	\If{$\overline{\Delta}_{l}=0$}{
	Apply $OP^l_{move}(l, \mathcal{N}, \mathcal{I})$;\\
	Remove $l$ from ${\mathcal{P}}$;\\
	}\Else{
	Run Subroutine $GA(l, \overline{\Delta}_l)$;\\
	If a neighbour $j$ of $l$, i.e., $j \in \partial l$, is in ${\mathcal{P}}$ remove $j$ from ${\mathcal{P}}$;\\
	$\forall j \in \partial l$, label each $j$ with the letter $C$;\\
	Build or update the $virtual$ node $\tilde{v}$ and, if it is not present, insert it into ${\mathcal{A}}$;\\
	Remove $l$ from ${\mathcal{P}}$;\\
	}
	}
}\uElseIf{${\mathcal{A}} \neq \emptyset$}{
	 \While{$\exists \tilde{v}\in {\mathcal{A}}$ s.t. $\overline{\Delta}_{\tilde{v}}=0$}{
	 Apply $SWAP-OP(\tilde{v})$;\\
	 Apply $OP^{\tilde{v}}_{move}(\tilde{v},  {\mathcal{A}})$;\\
	 } 
	 \While{$\exists \tilde{v}\in {\mathcal{A}}$ s.t. $\overline{\Delta}_{\tilde{v}}=1$}{
	 Apply $SWAP-OP(\tilde{v})$;\\
	 For $i\in \tilde{v}$ labeled $P$ s.t. $\overline{\Delta}_{i}=1$ run Subroutine $GA(i, \overline{\Delta}_i)$;\\
	 Pick $j \in \partial i$,with $j$ the last neighbour of $i$ added;\\
	Run Subroutine  $GA(j, \overline{\Delta}_j)$;\\
	 Apply $OP^j_{move}(j, \mathcal{N}, \mathcal{V})$;\\
	 Apply $OP^{\tilde{v}}_{del}(\tilde{v},  {\mathcal{A}})$;\\
	 }
	  \While{$\exists \tilde{v}\in {\mathcal{A}}$ s.t. $\overline{\Delta}_{\tilde{v}}=2$}{
	 Apply $SWAP-OP(\tilde{v})$;\\
	 $\forall i\in \tilde{v}$ labeled $P$ s.t. $\overline{\Delta}_{i} \leq 2$ run Subroutine $GA(i, \overline{\Delta}_i)$ and label the neighbour(s) of $i$ with the letter $C$;\\
	 Update the $virtual$ node $\tilde{v}$;\\
	 }
	Pick $\tilde{v}$ s.t. $\max_{\tilde{v}\in {\mathcal{A}}} \overline{\Delta}_{\tilde{v}}$;\\
	$\forall i\in \tilde{v}$ s.t. $\overline{\Delta}_{i} \neq 0$ and labeled $C$,  run Subroutine $GA(i, \overline{\Delta}_i)$;\\
	Apply $OP^{\tilde{v}}_{del}(\tilde{v},  {\mathcal{A}})$;\\ 
}\Else{
	 Pick a random site $m \in \mathcal{N}$ s.t. $\min_{m \in \mathcal{N}} \overline{\Delta}_m$ (not labeled $P$,or $C$);\\
	Apply $OP^{m}_{build-del}(m, \mathcal{N}, \mathcal{I}, \mathcal{V})$;\\
	 }
}
 \Return $\mathcal{I}$;
 \caption{local algorithm for $d>3$}
\end{algorithm}

\section{Conclusion}
\label{sec:num_anal}

This manuscript presents a new local prioritized algorithm for finding a large independent set in a random $d$-regular graph at fixed connectivity. This algorithm makes deferred decision in choosing which site must be set into the independent set or into the vertex cover set. This deferred strategy can be seen as a depth-first search delayed in time, without backtracking. It works, and shows very interesting results.

For all $ d \in [5, 100]$ we conjecture new lower bounds for this problem. All the new bounds improve upon the best previous bounds. All of them have been obtained by extrapolation on samples of random $d$-regular graphs of sizes up to $10^9$. For random $3$-regular graphs, our algorithm is able to reach, when $N \to \infty$, the asymptotic value presented in \cite{csoka2016independent}. For $4$-regular graphs, instead, we are not able to improve the existing lower bound. This discrepancy could be described by the fact that our algorithm is general and is not implemented only for a single value of $d$ with \textit{ad hoc} strategy. 

The improvements upon the best bounds are due to reducing the density of the graph, introducing regions in which virtual sites replace multiple original nodes and optimal labellings can be identified.
The creation of virtual sites allows to group together nodes of the graph to label at a different instant with respect to their creation. Those blobs of nodes transform the random $d$-regular graphs into a sparse graph, where the searching of a large independent set is simpler.

Undoubtedly more complex virtual nodes can be defined and additional optimizations can be identified.  This will be addressed in a future manuscript.

\paragraph{Acknowledgments.} R. M. would like to thank Nicolas Macris for a first reading of the manuscript, and Endre Cs\'oka for useful discussions. S. K. is supported by the Federman  Cyber Security Center at the Hebrew University of Jerusalem.  R.M. started this project when he was supported by the Federman  Cyber Security Center at the Hebrew University of Jerusalem, and finished it by the support of Swiss National Foundation grant No. 200021E 17554. 

\bibliographystyle{unsrt}  
\bibliography{references.bib}

\end{document}